\documentstyle[sprocl,epsfig]{article}

\def\be{\begin{equation}}
\def\ee{\end{equation}}
\def\bea{\begin{eqnarray}}
\def\eea{\end{eqnarray}}

\begin{document}

\title{ASYMPTOTIC GLUON SHADOWING}

\author{S. LIUTI}

\address{Institute of Nuclear and Particle Physics, University
of Virginia, \\ McCormick Road, Charlottesville, Virginia 22901, USA \\ and \\
INFN-Sezione Roma Tre, Dipartimento
di Fisica E. Amaldi, \\ Via Vasca Navale, 84. 00146 Roma, Italy.
\\E-mail: sl4y@virginia.edu } 

\author{F. CANO}

\address{Dipartmento di Fisica, Universit\'a di Trento,\\
O38050 Povo, Trento, Italy\\E-mail: cano@science.unitn.it}


\maketitle\abstracts{We examine the low Bjorken $x$ gluon distribution 
in nuclei in the asymptotic region. }

\section{Introduction}
Nuclear shadowing or the depletion at low Bjorken $x$
of the nuclear Deep Inelastic (DI) structure 
function, $F_2^A$, with respect to the nucleon one, $F_2^N$, has been 
observed in a number of experiments (see \cite{Arneodo} and references therein). 
Assuming the universality of parton distributions in nuclei, 
one expects nuclear shadowing to be present in other high energy processes as well, 
such as Drell-Yan pair, $J/\psi$ and $\Upsilon$ 
production in lepton-nucleus, hadron-nucleus and nucleus-nucleus 
collisions. 
In particular nuclear shadowing might be concurring with the other 
mechanisms among which quark-gluon plasma formation, 
that result in a depletion of the 
observed cross sections for these processes \cite{Gyu,Ramona}.
A quantitative understanding of both the $x$ and $Q^2$ dependences of the
nuclear parton distributions at low and very low $x$ is 
therefore a necessary step for interpreting  
the outcome of future experiments at RHIC and at the LHC. 
Recent calculations rely on   
non-perturbative models for 
the nuclear parton distributions at a given (low) scale, 
$Q_o^2$, combined with DGLAP \cite{DGLAP} perturbative evolution.
They are all therefore affected by the uncertainty in the initial parton
distributions and, in particular, in the gluon distribution which governs
evolution at low $x$, and which is poorly known experimentally. 
Moreover, evolution to moderate values of $Q^2$ 
($Q^2 \approx 10 \, {\rm GeV^2}$) depends 
even more strongly, by the value of $Q_o^2$ itself which,
if varied within the range, $Q_o^2 = 0.8 - few \, {\rm GeV^2}$
can lead to completely different evolution patterns
for the shadowing of both the 
structure functions and the gluon distribution.  
These simple facts hamper the possibility of predicting 
the $Q^2$ dependence of 
nuclear shadowing although its behavior within pQCD is quite well 
understood and it can be predicted
accurately for each choice of initial conditions. 
 
In this paper we suggest    
an avenue for dealing with the model dependence associated 
with the initial conditions of perturbative evolution, which is based on  
a study of the low $x$ 
and high $Q^2$ asymptotic behavior of the shadowing ratios $R_G=G_A/G_N$ and
$R_F=F_2^A/F_2^N$, $G_{N(A)}$ and $F_2^{N(A)}$ being the gluon distributions
and the Deep Inelastic (DI) structure function in a nucleon, $(N)$, and 
in an isoscalar nucleus, $(A)$, respectively. 
In the asymptotic regime defined by $Q^2 \gg Q_o^2$ and $x \ll x_o$, 
$x_o \leq 0.1$,
the 
Double Logarithmic Approximation (DLA) 
to pQCD evolution applies \cite{derujulaetal}.
The proton structure function data analyzed recently  
at HERA \cite{H1,ZEUS} have been proven \cite{H1} after the suggestion of 
Ref.\cite{BF_1}, to evolve according to DLA. 
The key test is to show that the data obey Double Asymtptotic Scaling (DAS) 
in the variables 
$\rho = \gamma ((Y-Y_o)/\xi)^{1/2}$ and 
$\sigma = \gamma^{-1}((Y-Y_o)\xi)^{1/2}$, $\gamma=6/(33-2N_f)^{1/2}$, 
$Y=\ln 1/x$, 
$\xi=\gamma^2\ln (\ln Q^2/\Lambda_{QCD}^2/\ln Q_o^2/\Lambda_{QCD}^2$).
Violations from DAS (other than due to the fact that the data lie in a 
pre-asymptotic region or to NLO DGLAP corrections \cite{BF_1}b) 
would signal either the onset of other approximations such as 
leading-(next-to-leading)-log in $1/x$, $LL_x$ ($NL_x$) resummation, 
or the beginning of parton saturation. 
Here we wish to apply a DAS type of analysis to nuclear DI scattering. 

As a first step we show that if evolution proceeds through 
ordinary asymptotic pQCD evolution equations in nuclei as well as 
in a free nucleon, 
the ratios $R_F$ and $R_G$ become a function of $\rho$ only, which is {\it per se} 
a model independent result. We use this result as a basis for exploring 
the origin of scaling violations in nuclei. 
In nuclei in fact, the asymptotic regime is entered in principle at different
values of $x_o^A$ and $Q_{o,A}^2$ than in the proton.   
In particular Unitarity Shadowing  
Corrections (USC) \cite{GLR,MueQiu,CK,EQW} 
are expected to affect evolution at  
$x_o^A > x_o$ because of the increase of the gluon density in a nucleus
due to the overlapping of nucleons in the longitudinal direction. 
On a more speculative basis one might also expect the transition to the 
$\ln(1/x)$ resummation to appear in 
a different regime. 
In our approach such questions can be addressed systematically as 
they introduce
specific {\em scaling violations} from the DLA result, 
appearing as a $\sigma$ dependence in the ratios $R_G$ and $R_F$.

In summary, although it is technically 
predictable that in a proton 
at very low values of $x$ and sufficiently large $Q^2$, {\it i.e.} 
deeply in the asymptotic region, DGLAP evolution  
and the DLA should break down
and give way to $\ln (1/x)$ summation and to USC,    
it is still a major task to be able to pinpoint where and if the 
transition from the different regimes is going to take place
in the kinematical regimes currently under exploration.
Our goal is to obtain some new insight  
by using nuclear targets where the asymptotic regime
can in principle be reached at larger $x$. 
As a by-product we obtain quantitative predictions for RHIC and the LHC. 

\section{DGLAP Evolution in Nuclei}
We first summarize results for ordinary DGLAP evolution applied to the nuclear
ratios at low $x$, assuming that the proton and the nuclear distributions evolve
along similar paths. 
As it is well known evolution is driven by the gluon distribution 
which dominates over the
sea quarks one and  
one can predict the behavior of the shadowing 
ratios, $R_G$ and 
$R_F$ with $Q^2$:
\begin{eqnarray}
\frac{\partial R_G }{\partial \ln Q^2} \simeq  
 \int_x^1     
P_{GG}\left(\frac{x}{y},\alpha_S(Q^2)\right) \frac{G_N(y,Q^2)}{G_N(x,Q^2)}
& &
\nonumber \\ 
\times  \left[ R_G(y,Q^2)- R_G(x,Q^2)\right] \frac{dy}{y} 
& & \nonumber \\
\propto  \frac{\partial G_A(x,Q^2)/\partial \ln Q^2}
                 {\partial G_N(x,Q^2)/\partial \ln Q^2} - R_G(x,Q^2) & &, 
\\
\frac{\partial R_F}{\partial \ln Q^2}  \simeq 
\int_x^1  
P_{qG}\left(\frac{x}{y},\alpha_S(Q^2)\right) 
\frac{G_N(y,Q^2)}{\Sigma_N(x,Q^2)} & & \nonumber  \\ 
 \times \left[ R_G(y,Q^2)- R_F(x,Q^2)\right] \frac{dy}{y} & &,   
\end{eqnarray} 
\noindent 
where we have disregarded the sea quarks distribution on the 
{\it r.h.s.} of the coupled DGLAP evolution equations; 
$P_{qG}$ and $P_{GG}$ are the splitting functions 
evaluated at NLO; and we used the approximation 
$F_2^{N(A)} \approx 5/18 \Sigma_{N(A)}$.
Eqs.(1) and (2) show that the $Q^2$ dependence of $R_G$ and $R_F$ is 
regulated by a subtle balance involving the parton distributions 
and the ratios themselves \cite{Qiu}. 
The following predictions can be made for the
ratios $R_G$ and $R_F$: If, as predicted by current non-perturbative shadowing 
models
$R_G$ is a growing function of $x$, then it also grows with $Q^2$, the {\it r.h.s.}
of Eq.(1) being positive ($y \geq x$). 
On the other side, defining
$R_G(x,Q_o^2) \equiv R_G^o$ and $R_F(x,Q_o^2) \equiv R_F^o$, one obtains 
two opposite behaviors for $R_F$ namely: {\em i)} if $R_F^o < R_G^o$, 
then $R_F$ grows with $Q^2$; {\em ii)} if instead 
$R_F^o > R_G^o$, then $R_F$ initially decreases with $Q^2$ until 
it reaches the value of $R_G^o$ and it subsequently starts increasing 
along with $R_G$.  
\footnote{Although the form of Eq.(1) might seem 
suggestive of a fixed point behavior \cite{HLS}, 
this is {\it a priori} not the case, since 
the quantity $\partial G_A(x,Q^2)/ \partial \ln Q^2 / \partial G_N(x,Q^2)/
\partial \ln Q^2$ depends on $Q^2$.}
The ``rate'' of change with $Q^2$ is 
governed by the ratios $G_N(y)/G_N(x)$ and $G_N(y)/\Sigma_N(x)$, at $Q^2=Q_o^2$,
respectively. If $Q_o^2 \leq 1 \, {\rm GeV^2}$, then a rapid 
evolution strongly reduces the shadowing in both $R_G$ and $R_F$, 
between $Q_o^2$ and $Q^2 \approx 2-3 \, {\rm GeV^2}$.  
If on the contrary $Q_o^2$ ranges from $2$ to $5 \, {\rm GeV^2}$ then the evolution 
is slower and shadowing remains large even at $Q^2 \approx 10-100 \, {\rm GeV^2}$
(these results affect quarkonia production where $Q^2 \equiv M^2_{J/\psi} 
\approx 10 \, {\rm GeV^2}$ and $Q^2 \equiv M^2_{\Upsilon} \approx 100 \, 
{\rm GeV^2}$). 


\section{Double Asymptotic Scaling}
We now examine the nuclear DI structure function and gluon distribution 
in the asymptotic regime.
The derivation of the equations of the DLA in a nucleus parallels 
the one for the proton, namely one first writes   
the LO DGLAP evolution equation for the gluon distribution in the limit 
$x \rightarrow 0$, in moments space (we have omitted the subscripts $N(A)$
unless necessary):
\begin{equation}
\frac{\partial g(n,Q^2)}{\partial \ln Q^2}   =  \frac{\alpha_s(Q^2)}{2 \pi} 
\gamma^0_{GG}(n) g(n,Q^2), 
\end{equation}
$\gamma^0_{gg}(n) \approx 2 C_A/(n-1)$ being the anomalous dimension in 
the limit $n \rightarrow 1$. 
Solutions in $(x,Q^2)$ are found by evaluating the anti-Mellin transform, 
\begin{eqnarray}
G(x,Q^2) & = & \frac{1}{2 \pi i} \int_C dn \, g(n,Q_0^2) \nonumber \\ 
& & \exp\left\{ (n-1)Y + \xi/(n-1) \right\},  
\label{DLLA0}
\end{eqnarray}
with the saddle point method. In Eq.(\ref{DLLA0})
$g(n,Q_0^2) = \int_0^1 dx \, x^{(n-1)} g(x,Q_0^2)$, $g(x,Q_o^2)$ being the
inital gluon distribution, and $G(x,Q^2)=xg(x,Q^2)$.   
If one takes a ``soft'' initial condition such as, 
$G(x,Q_0^2) \approx A_N x^{-\lambda}$, 
$\lambda \approx 0$, and $Y$ and $\xi$ are both similarly large, then
$g(n_0,Q_0^2) \approx A_N/(n-1)$, and the saddle point is:
$n_0= 1/(2\Delta Y) + \sqrt{1/(4\Delta Y)^2+\xi/\Delta Y} \approx 
\sqrt{1+\xi/ \Delta Y}$, $\Delta Y = Y- Y_o$, $Y_o=\ln(1/x_o)$, yielding:
\begin{equation}
G(x,Q^2)  =  \sqrt{2 \pi}  
\left( \frac{\widetilde{g}(n_o,Q_o^2)}
{\widetilde{g}''(n_o,Q_o^2)} \right)^{1/2}
\widetilde{g}(n_o,Q^2),
 \label{DLLA1}
\end{equation}
with $\widetilde{g}(n,Q^2)=g(n,Q^2)\exp[(n-1)Y]$.
\footnote{We have omitted sub-leading corrections for simplicity.}
In terms of the DAS variables, $\rho$ 
and $\sigma$, $n_0= 1 + \rho/\gamma^2$
and:
\begin{equation}
G \equiv G^{DLA}(x,Q^2) = f(\rho,\sigma) \exp(2\gamma\sigma),
\label{DLLA2}
\end{equation}
where $f(\rho,\sigma)$ is a function that depends on the 
explicit form of the initial
conditions.
The main result is that $\ln(G^{DLA})/f(\rho,\sigma))$ is predicted to be 
linear in $\sigma$, the slope being fixed by 
$\gamma$. 
A similar behavior is found for $F_2^N$ by solving the equation:
\begin{equation}
\frac{\partial F_2^N(x,Q^2)}{\partial \ln Q^2} = 
\frac{2}{9} \frac{\alpha_S}{\pi} G(x,Q^2)
\label{F2DLA} .
\end{equation}
Deviations from DAS due to hard initial 
conditions
{\it i.e.} when $\lambda \geq 0.2$ 
are technically obtained as follows: 
$g(n_0,Q_0^2) \approx A_N/(n-(\lambda+1))$
and a physical solution in the limit $Y \gg \xi$ is
\begin{equation}
G(x,Q^2) = C/\lambda e^{\lambda (Y-Y_o) + \xi/\lambda},  
\label{HardP}
\end{equation}
corresponding to the saddle point value,
$n_0(\lambda)= 1/\Delta Y + \lambda + 1$ (see also \cite{BF_1,CK}).
Physically this behavior supports the appearance of USC
which become more important just because of the steep rise at small $x$.
The data of Ref.\cite{H1} favor the DAS scenario,
whereas the more recent data at lower $x$ and $Q^2$ \cite{ZEUS} 
are best interpreted in terms of USC \cite{Levinetal}.
In this context it is therefore important to study nuclear shadowing 
because, due to the $A$-dependent increase in gluon density,  
the physics we are interested in is expected to show up at larger 
$x$ values and because one can tune the variable $A$ 
in order to discriminate among models.

In a nucleus we make the two following assumptions for the 
asymptotic regime: {\it i)} the initial distributions are shadowed because 
of some non-perturbative mechanism, the evolution equations are not
affected by the medium; {\it ii)} besides the non-perturbative shadowing 
the evolution equations have screening corrections at a larger $x_A$. 
The derivation of the ratio $R_G$ in the hypothesis {\it i)} yields:
\begin{eqnarray}
R_G^{Asym} = \left[\frac{G_A(n_o,Q^2)}{G_N(n_o,Q^2)} \right]^{3/2}
\left[\frac{G_N''(n_o,Q^2)}{G_A''(n_o,Q^2)} \right]^{1/2} & & \nonumber 
\\ =
f_A(\rho) \frac{\rho h^{(1)}_N(\rho) + \gamma^2 \sigma h^{(2)}_N(\rho)}
{\rho h^{(1)}_A(\rho) + \gamma^2 \sigma h^{(2)}_A(\rho)} & &.
\label{RGAsym}
\end{eqnarray} 
where $f_A(\rho)=G_A(n_o,Q_o^2)/G_N(n_o,Q_o^2)$, and the functions
$h_{N(A)}^{(1,2)}$ are listed explicitely elsewhere. The main point  
is that the exponential terms appearing in $G^{DLA}$, Eq.(\ref{DLLA2}) 
calculated for a nucleus and for a nucleon respectively,   
cancel exactly and the remaining dependence 
on $\sigma$ is small asymptotically.   
This scaling result (shown in Fig.1) will be our reference point. 
\begin{figure}[t]
\vbox{
\hskip.6truecm\epsfig{figure=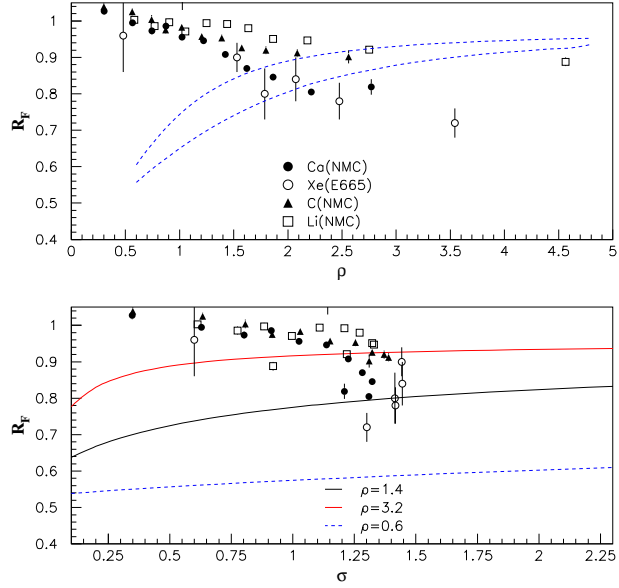,width=9.truecm}}
\vskip-.6truecm
\caption{Nuclear structure functions ratios, $R_F$, vs. 
the DAS variables $\rho$ (top) and $\sigma$ (bottom). 
The theoretical curves
\protect{\cite{CanLiu}} show the onset of scaling in $\sigma$. The  
non-perturbative shadowing model used in the calculation is from 
Ref.\protect{\cite{FS}}. Experimental
data from \protect{\cite{NMC1,NMC2}} (Li, C, Ca) and 
\protect{\cite{E665}} (Xe).}
\label{fig1}
\end{figure}

The onset of a different evolution 
mechanism in the nucleus will appear as a $\sigma$-{\em scaling violation} 
modifying the exponential behavior of Eq.(\ref{DLLA1}). In 
general one can write
\begin{equation}
R_G = 
R_G^{Asym}(\rho) \times \exp \left\{2\gamma (\sigma_A-\sigma) \right\},
\end{equation}
where $R_G^{Asym}$ is the large $\sigma$ limit of Eq.(\ref{RGAsym}).
The form of $\sigma_A\equiv \sigma_A(x,Q^2)$ describes 
the approximation to pQCD evolution in a nucleus. In the case 
of a ``hard pomeron'' boundary conditions 
plus screening corrections \cite{GLR} it is
\begin{equation}
\sigma_A = D_A(x,Q^2) \left( \lambda \sigma \rho + 
\frac{\gamma^2}{\lambda} \frac{\sigma}{\rho} \right),  
\end{equation} 
$D_A$ being a damping correction 
(more details are going to be found in \cite{CanLiu}).  

In conclusion, we have shown model independent predictions
for gluon shadowing in nuclei in the asymptotic region. 
We have also outlined an approach  
for more detailed studies of the violations from DAS in nuclei
as a signal of pQCD evolution mechanisms other than DGLAP or the
DLA.
Our calculations will be relevant in the regime accessible
at future experiments at RHIC, LHC
and possibly at the $eA$ project at DESY \cite{eA}.

\section*{Acknowledgments}
The authors wish to thank the Institute of Nuclear and Particle Physics at the 
University of Virginia for financial support.

\end{document}